\documentclass{article}
\usepackage{geometry}                
\usepackage{graphicx}
\usepackage{stmaryrd}
\usepackage{amssymb}
\usepackage{amsthm, bm}
\usepackage{bbm}
\usepackage{amsmath, cancel, centernot }
\usepackage[mathscr]{eucal}
\usepackage{mathtools}
\usepackage{epstopdf}
\usepackage{mathrsfs}
\usepackage{newtxmath}

\title{Impossible Counterfactuals, Discrete Hilbert Space and Bell's Theorem}
\author{Tim Palmer\\ Department of Physics, University of Oxford, UK\\
tim.palmer@physics.ox.ac.uk}
\date{20 January 2026}
\makeatletter
\newcommand\be{\@ifstar{\[}{\begin{equation}}}
\newcommand\ee{\@ifstar{\]}{\end{equation}}}
\newcommand\bp{\begin{pmatrix}}
\newcommand\ep{\end{pmatrix}}

\newtheorem*{theorem*}{Theorem}

\makeatother
\begin{document}
\bibliographystyle{plain}
\maketitle
{\centering \emph{In fond memory of Basil Hiley.} \par}

\abstract{Negating the Measurement Independence assumption (MI) is often referred to as the `third way' to account for the experimental violation of Bell's inequality. However, this route is generally viewed as ludicrously contrived, implying some implausible conspiracy where experimenters are denied the freedom to choose measurement settings as they like. Here, a locally realistic model of quantum physics is developed (Rational Mechanics - RaQM - based on a gravitational discretisation of Hilbert Space) which violates MI without denying free will. Crucially, RaQM distinguishes experimenters' ability to freely choose measurement settings to some nominal accuracy, from an inability to choose exact settings, which were never under their control anyway. In RaQM, Hilbert states are necessarily undefined in bases where squared amplitudes and/or complex phases are irrational numbers. Such `irrational' bases correspond to conceivable but necessarily impossible counterfactual measurements, and are shown to play a ubiquitous role in the analysis of both single- and entangled-particle quantum physics. It is concluded that violation of Bell inequalities can be understood with none of the strange processes historically associated with it. Instead, using concepts from (non-classical) $p$-adic number theory, we relate RaQM to Bohm and Hiley's concept of a holistic Machian-like Undivided Universe. If this interpretation of Bell's Theorem is correct, building more and more energetic particle accelerators to probe smaller and smaller scales, in the search for a theory which synthesises quantum and gravitational physics and hence a Theory of Everything, may be a fruitless exercise.}

\section{Introduction}

I first met Basil Hiley as a result of a complete misunderstanding. One day in the early 1990s, I was idly perusing the physics books in a well-known Oxford book store. By chance, I came across the volume \cite{HileyPeat} edited by Basil Hiley and David Peat entitled: `Quantum Implications: Essays in Honour of David Bohm'. On the front cover was a picture of a fractal. I was excited and wrote to Basil asking for more details about the relevance of this front cover image, explaining that, according to my way of thinking \cite{Palmer:1995}, if the quantum potential of Bohmian theory could be treated as a coarse-grain approximation to some fine-scale fractal state-space geometry, it may be possible for Bohmian theory to violate Bell's inequality without having to invoke EPR/Bell nonlocality. Basil wrote back and said that unfortunately the picture on the cover had been chosen by the publisher for no other reason than it looked sexy!  However, Basil added that he found my idea intriguing and would be interested to learn more. That started a number of journeys to Sweden to participate in some fascinating summer workshops on quantum foundations organised by Basil and Georg Wikman. 

In this paper I will describe why I felt then, and do so even more today, that the fractal on the cover of Basil and David's book is (serendipitously) central to understanding the deep non-classical properties of quantum physics. To get to this, I describe in Section \ref{SCD} my view that Bell's Theorem should be understood as a constraint on the notion of Counterfactual Definiteness (CD) in quantum (but not classical) physics. Such a view is idiosyncratic, and over the years I have encountered two arguments against this perspective. The first is that CD is an irrelevance for Bell's Theorem. The second is that if we can't rely on CD, then we can't assume that experimenters have free will, since we can no longer assert that they could have done otherwise.  And if experimenters don't have free will, then it makes, in the words of Anton Zeilinger, \cite{Zeilinger} 'no sense at all to ask nature questions in an experiment, since then nature could determine what our questions are, and that could guide our questions such that we arrive at a false picture of nature'. To the first group, I agree that most discussions of Bell's Theorem do not single out CD as an assumption. However, as discussed in Section \ref{SCD}, a number of physicists and philosophers (including a Nobel Laureate) have drawn specific attention to CD, as being implicit in the Measurement Independence (MI) assumption. The bulk of this paper is instead addressed to the second group. In this paper I will be discussing a model of quantum physics (based on the Schr\"{o}dinger equation) where experimenters have complete freedom to choose measurement settings as they like. However - and this is the key `new meat' of the paper - when they choose such settings then, inevitably, they can only choose them to some nominal accuracy. For example, if an experimenter performs a measurement on a particle using a Mach-Zehnder interferometer, they can freely choose the arms of the interferometer to have equal length, but only to some nominal accuracy. They manifestly cannot choose the arms to be precisely equal in length since the experimenter has no control over whether a gravitational wave passes by at the time of measurement. As will be discussed, incorporating this distinction between nominal and exact measurement settings into the proposed model, makes all the difference between a theory with on the one hand a plausible, and on the other hand a grotesquely conspiratorial violation of MI. This distinction is reminiscent of the role of finite precision measurements in relation to the Kochen-Specker Theorem \cite{Meyer:1999}.

The model of quantum physics, discussed in Section \ref{discrete}, is called Rational Quantum Mechanics (RaQM) - a full exposition of RaQM will be given elsewhere. In RaQM, as in quantum mechanics (QM) itself, Hilbert states are evolved using the Schr\"{o}dinger equation without modification. However, in RaQM, unlike in QM, such states are only mathematically defined in bases where squared amplitudes and complex phases (in degrees) are rational numbers. In particular, Hilbert states are undefined in bases when squared amplitudes and/or complex phases are irrational numbers. Such `irrational' states correspond to conceivable but impossible counterfactual worlds. It turns out that these states play a ubiquitous role in characterising the properties of the quantum world. As a warm up to Bell's Theorem, in Section \ref{complementarity} we discuss the failure of CD as a way to understand complementarity and non-commutativity of single-particle observables. In Section \ref{2adic} we discuss whether a theory which eschews irrational numbers is necessarily fine-tuned. To show that it is not, we refer to the two metrics from which one can complete the set $\mathbb Q$ of rational numbers. We argue that whilst the Euclidean metric is the relevant metric to measure distances between objects in physical space, the (non-classical) $p$-adic metric is the relevant metric to measure distances between states of the world in state space. 

Using these ideas, Bell's Theorem is discussed in Section \ref{Bell} using this novel approach to MI. A key conclusion of this paper is that entanglement does not imply any of the `weird' things that have historically been attributed to it: EPR/Bell nonlocality, an unbounded speed limit for quantum information, non-realism, conspiracy, retrocausality, wormholes, branching of worlds, and so on. Rather, like Mach's Principle, Bell's Theorem is telling us that the laws of physics are holistic. In Section \ref{conc}, using concepts from $p$-adic fractal geometry, we discuss RaQM in relation to the author's holistic Invariant Set Postulate \cite{Palmer:2009a} and Bohm and Hiley's related concept of an Undivided Universe \cite{BohmHiley}. 

\section{Measurement Independence and  Counterfactual Definiteness}
\label{SCD}

In discussing Bell's Theorem in the early 1970s the American physicist Henry Stapp \cite{Stapp} wrote:
\begin{quote}
The third way out [other than violating locality and realism] is to deny that the measurements that `could have been performed, but were not', would have had definite results if they had been performed.
\end{quote}
Nobel Laureate Anton Zeilinger described the same idea is his popular account of Bell's Theorem \cite{Zeilinger}:
\begin{quote}
There is a third assumption $\ldots$. It is the assumption that it makes sense to consider what kind of experimental results would have been obtained if one had measured a different property than the one that was actually measured. 
\end{quote}
In his monograph, Michael Redhead made much the same inference when he wrote about Bell's Theorem
\begin{quote}
A very important point to notice here is that the possessed values [of spin in Bell's inequality] are in general counterfactually possessed. 
\end{quote}
These quotes are specific interpretations of the more mathematically precise MI assumption 
\be
\label{MI}
\rho(\lambda | \boldsymbol A, \boldsymbol B)=\rho(\lambda),
\ee
where $\lambda$ is a hidden variable and $\boldsymbol A$, $\boldsymbol B$ denote Alice and Bob's measurement orientations in a run of a Bell experiment and $\rho$ is a probability function. To begin to understand Stapp, Zeilinger and Redhead's appeal to a possible violation of CD, suppose each entangled particle pair is described by a unique $\lambda$. Then in the real world in which a Bell experiment is performed, each $\lambda$ is associated with some specific pair of real-world measurement orientations (one for Alice and one for Bob). If these settings correspond to $\boldsymbol A$ and $\boldsymbol B$, then any other pair of measurement settings such as $\boldsymbol A$ and $\boldsymbol C$, or $\boldsymbol B$ and $\boldsymbol C$, for that specific $\lambda$, are necessarily counterfactual. The question then arises: can one envisage a consistent and plausible theoretical model of quantum physics where measurement outcomes associated with such counterfactual worlds might be undefined?

It is typically believed that the answer to this question is no. Indeed, (\ref{MI}) is often described as the Free Choice assumption \cite{Blasiak} - experimenters are free to set their measuring apparatuses as they like. Free choice - that one could have chosen otherwise for a given $\lambda$ - implies that $\rho(\lambda)$ should not depend on such measurement settings. One could perhaps imagine some grotesquely contrived conspiracy where experimenters' minds are somehow subverted to only choose settings that will guarantee measurement outcomes consistent with QM. However, as has been written many times, surely no sane physicists would countenance such a bizarre state of affairs \cite{Aaronson}. 

However, as discussed in the Introduction, and seemingly uniquely in the vast literature on the subject, here we make a fundamental distinction between nominal and exact measurement settings. In the model of quantum physics described in Section \ref{discrete}, experimenters are free to choose measurement settings (e.g. polariser orientations $\boldsymbol A_{\mathrm{nom}}$, $\boldsymbol B_{\mathrm{nom}}$) to some nominal accuracy, but are not free to choose exact measurement settings (e.g. exact polariser settings, for which we use the symbols $\boldsymbol A$, $\boldsymbol B$ without subscript). For example, in considering the phase difference $\phi$ between the two arms of a Mach-Zehder interferometer, an experimenter is free to choose a nominal value $\phi_{\mathrm{nom}}=0$, but manifestly has no control on the effect of a gravitational wave, emitted from the collision of two distant black holes, on the precise value of $\phi$. In the discussions below, we will be pursuing a theory of quantum physics based on discretised Hilbert Space, where, as a consequence of discretisation
\begin{align}
\label{MI2}
\rho(\lambda | \boldsymbol A_{\mathrm{nom}}, \boldsymbol B_{\mathrm{nom}})&=\rho(\lambda) \nonumber \\
\rho(\lambda | \boldsymbol A, \boldsymbol B)&\ne \rho(\lambda)
\end{align}
As will be discussed in Section \ref{2adic}, making such a distinction between nominal and exact settings does make robust physical sense if the metric on state space is $p$-adic rather than Euclidean. As will be shown, (\ref{MI2}) provides a novel interpretation of the experimental violation of Bell's inequality, without needing to invoke nonlocality, conspiracy, violation of free choice, wormholes in space-time, Everettian branching, or any of the other strange ideas that have been mooted over the years to explain the violation of Bell's inequality. By way of support for this conclusion, finite measurement precision is known to play a role in the interpretation of the Kochen-Specker Theorem \cite{Meyer:1999}. 

A theory which violates MI is sometimes referred to as `superdeterministic'. Typically, such a theory is rejected on the basis that such violation implies some implausible conspiracy where experimenters' choices of measurement setting and particle hidden variables are somehow correlated. It is important to note that no such conspiracies are implied by the failure (\ref{MI2}) of MI for exact settings. As discussed, the experimenters are completely free to choose measurement settings to any finite nominal accuracy. Insofar as superdeterminism describes any theory which violates MI (including (\ref{MI2}), then RaQM is superdeterministic. However, if the word superdeterministic is to be reserved for theories which invoke an implausible correlation between measurement settings and hidden variables, then RaQM is not superdeterministic. 

\section{Discretised Hilbert Space - Rational Quantum Mechanics}
\label{discrete}
\subsection{Preamble}

There are 3 compelling reasons for discretising Hilbert Space, some of which are discussed in this paper, others in the companion papers \cite{Palmer:2025a}:
\begin{itemize}
\item It helps solve some of the long-standing conceptual problems that have dogged QM since its inception, This is the focus of this paper. 
\item The eminent physicist John Wheeler who famously invented the aphorism `It from Bit' (reality from information) wrote \cite{Wheeler}
\begin{quote}
The familiar probability function or functional, and wave equation or functional wave equation, of standard quantum theory provide mere continuum idealizations and by reason of this circumstance conceal the information-theoretic source from which they derive.
\end{quote}
Hilbert Space discretisation reveals explicitly the information-theoretic nature of the quantum state, concealed by QM. 
\item The measurement problem can be solved. Without adding terms to the Schr\"{o}dinger equation, quantum state reduction and measurement can be described  in two equivalent ways: in terms of a loss of information and in terms of a chaotic evolution to stable attractors.
\end{itemize}
In RaQM, the physical basis for Hilbert Space discretisation is gravity; the fact that gravity is weak implies that discretisation is very fine indeed (and thus difficult to detect). QM is the singular continuum limit of discrete Hilbert Space where the gravitational constant $G$ in RaQM is set to zero precisely. The results discussed in this paper hold no matter how fine the discretisation is, as long as it is non-zero. The author has proposed an experiment, potentially achievable within 5 years, which can test RaQM against QM \cite{Palmer:2025a} \cite{Palmer:1995}

\subsection{Single-particle discretisation}

In RaQM, Schr\"{o}dinger evolution is not modified. Rather, we impose the restriction that a unitarily evolved qubit state 
\begin{equation}
\label{qubit3}
|\psi(\theta,\phi) \rangle= \cos \frac{\theta}{2} |1\rangle + e^{i \phi} \sin\frac{\theta}{2} |-1\rangle
\end{equation}
is only mathematically defined in bases $\{|1\rangle, |-1\rangle\}$ (eigenfunctions of some quantum observable) where
\begin{equation}
\label{rat}
\cos^2 \frac{\theta}{2}= \frac{m}{L} \in \mathbb Q ; \ \ \ \ \phi= 2\pi \frac{n}{L} \in \mathbb Q
\end{equation}
Here $L \in \mathbb N$ is a mass/energy-dependent variable which defines the degree of granularity of discretised Hilbert Space, and $0 \le m, n \le L$ are whole numbers. According to \cite{Palmer:2025a}, $L \approx 10^{100}$ for a typical qubit in, say, a quantum computer. Relative to a basis where the rationality conditions (\ref{rat}) do not hold (e.g. if squared amplitudes or phases in degrees are irrational), the Hilbert state (\ref{qubit3}) is undefined. 

As described in the Appendix, the key to RaQM's representation of the quantum state lies in the discretised representation of complex numbers and quaternions as permutation/negation operators acting on finite-length bit strings  (see also \cite{Palmer:2020} \cite{Palmer:2024} and a more comprehensive paper on RaQM in preparation). Using this discretisation of complex numbers, (\ref{qubit3}) can be represented as a length-$L$ bit string. Specifically, if $\theta$, $\phi$ satisfy  (\ref{rat}) then
\begin{align}
\label{bits}
|\psi(\theta, \phi) \equiv \{\underbrace{\ 1,\; \ \; 1,\ \; 1,\ \ \dots \ \ 1,\ \; 1,\ \; 1,\ \ -1,-1,-1, \ \ldots \ ,-1,-1,-1}_{\theta, \phi}\}\ \ \ \bmod \xi 
\end{align}
comprising the bits `1' and '-1' (symbolic labels for measurement outcomes \cite{Schwinger}). Here $\cos^2 \theta/2 = m/L$ equals the fraction of 1s in the bit string (hence $\sin^2 \theta_1/2$ equals the fraction of -1 bits), and $\phi/2\pi=n/L$ encodes $n$ cyclic permutations of the bit string, corresponding to a complex phase rotation in QM. The Hilbert state in QM is invariant under a global phase. Corresponding to this, the state in RaQM can be considered an equivalence class of bit strings under general permutations $\xi$. In this sense, (\ref{bits}) describes an unordered ensemble of $L$ possible binary measurement outcomes with frequencies consistent with Born's rule (in the rational basis with respect to which the quantum state is defined). 

However, unlike in QM, global phase plays a key role in RaQM. In particular, for some specific $\xi$ we write
\begin{align}
\label{bits2}
|\psi(\theta, \phi) \rangle_\xi = \xi  \{\underbrace{\ 1,\; \ \; 1,\ \; 1,\ \ \dots \ \ 1,\ \; 1,\ \; 1,\ \ -1,-1,-1, \ \ldots \ ,-1,-1,-1}_{\theta, \phi}\}
\end{align}
describing a specific ordered ensemble of bits. Indeed, going further, we can map (\ref{bits2}) to the binary representation of a single integer (by replacing each -1 with 0). This integer describes the state of \emph{a specific} quantum system e.g. a specific photon, in the rational basis. As such, $\xi$ can be thought of as a hidden variable (hidden to us), describing the relationship between the quantum system and the rest of the universe, and fixed the moment the quantum system is created. $\xi$ is key to describing the outcome $\pm 1$ of a measurement on a specific individual quantum system. The measurement outcome associated with a measurement on $|\psi\rangle_\xi$ corresponds to the first member of the bit string after the application of $\xi$. In \cite{Palmer:2025a}, this is seen as the result of a (chaotic) shift map repeatedly applied to the integer associated with $|\psi\rangle_\xi$. For the purposes of this paper, we can think of $\xi$ as fixing which of the $L$ bits in (\ref{bits2}) corresponds to the measurement outcome \emph{before} the application of $\xi$. We'll refer to this as the $1 \le M(\xi) \le L$th bit. Since $\xi$ is hidden, we can for all practical purposes think of the $M(\xi)$th bit as randomly chosen. However, it is important to note that the notion of randomness here is not fundamental, any more than rolling a dice is fundamentally random. Effectively, $\xi$ allows a deterministic description of single-particle states, not possible in QM. 

As discussed, a single quantum particle (in a specific rational basis) can be described by a length-$L$ bit string. $L$ is referred to as `qubit information capacity'; $L=\infty$ in QM. The finiteness of $L$ should be testable experimentally in a few years using quantum computers \cite{Palmer:2025a}. 

\subsection{Multiple Qubits and EPR/Bell Locality}

In RaQM, a quantum system comprising $N$ entangled qubits is represented as $N$ correlated length-$L$ bit strings, where the same permutation $\xi$ applies to each string (consistent with $\xi$ as a global phase in QM). As discussed in \cite{Palmer:2025a}, and providing $L$ is large enough, there are as many of degrees of freedom in these $N$ bit strings, as in an $N$ qubit system in QM ($2^{N+1}-2$). In this paper, we will only be concerned with $N=1$ or $N=2$. A general, explicitly normalised, $N=2$-qubit state in QM can be written
\begin{equation}
\label{threeraqm}
\begin{split}
\begin{array}{cc}
|\psi_{AB}\rangle=\underbrace{\cos \frac{\theta_1}{2}|1_A\rangle}_A\  
\times\  (\underbrace{\cos\frac{\theta_2}{2}|1_B\rangle+e^{i \phi_2} \sin\frac{\theta_2}{2}|-1_B\rangle}_B)+ \\
\ \ \ \ \ \ \ \ \ \ \  \ \  \ \ \  \ \ \  \ \  +\underbrace{e^{i \phi_2}\sin \frac{\theta_1}{2}|-1_A\rangle}_A
\times (\underbrace{\cos\frac{\theta_3}{2}|1_B\rangle+e^{i \phi_3} \sin\frac{\theta_3}{2}|-1_B\rangle}_B) 
\end{array}
\end{split}
\end{equation}
If the individual amplitudes and phases in (\ref{threeraqm}) satisfy (\ref{rat}), then in RaQM $|\psi_{AB}\rangle$ is equivalent to the 2 length-$L$ bit strings
\be
\label{twobitstrings}
|\psi_{AB}\rangle \equiv \left\{
\begin{array} {c}
\{\underbrace{\ 1,\; \ \; 1,\ \; 1,\ \ \dots \ \ 1,\ \; 1,\ \; 1,\ \ -1,-1,-1, \ \ldots \ ,-1,-1,-1}_{\theta_1, \phi_1}\}\ \ \ \bmod \xi  \\
\{\underbrace{1,1,\; \ldots , 1, \ -1, -1,\ldots,-1}_{\theta_2, \phi_2} \ \underbrace{\ 1,1, \ldots ,\ \ 1,\ -1,-1 \ldots \ ,-1}_{\theta_3, \phi_3}\}\ \ \ \bmod \xi 
\end{array}
\right.
\ee
with 6 degrees of freedom, as in QM. The top string represents the A qubit, and the bottom representing the B qubit, where for each $i$, $\theta_i$ and $\phi_i$ satisfy the rationality conditions (\ref{rat}). Hence, for example, $\cos^2 \theta_2/2$ denotes the fraction of $1$ bits in the B bit string which correspond to 1 bits in the A bit string, and $\phi_2$ represents a cyclic permutation of specific bits in the B bit string, as shown. Again, $\xi$ corresponds to a permutation common to both bit strings, corresponding to some global phase. As such, in RaQM, the quantum state for a specific pair of qubits can be written  
\be
\label{twobitsxi}
|\psi_{AB}\rangle_\xi=\left\{
\begin{array} {c}
\xi \{\underbrace{\ 1,\; \ \; 1,\ \; 1,\ \ \dots \ \ 1,\ \; 1,\ \; 1,\ \ -1,-1,-1, \ \ldots \ ,-1,-1,-1}_{\theta_1, \phi_1}\}  \\
\xi \{\underbrace{1,1,\; \ldots , 1, \ -1, -1,\ldots,-1}_{\theta_2, \phi_2} \ \underbrace{\ 1,1, \ldots ,\ \ 1,\ -1,-1 \ldots \ ,-1}_{\theta_3, \phi_3}\}
\end{array}
\right.
\ee

For Bell's Theorem, we focus on the QM singlet state
\be
\label{singlet}
|\psi^{\mathrm{singlet}}_{AB}\rangle=\frac{1}{\sqrt 2}\left (|\ 1_A\rangle |-1_B\rangle-|-1_A\rangle|\ 1_B\rangle \right)
 \ee
where A and B denote Alice and Bob's qubit respectively. With respect to a basis where Alice's measuring apparatus is oriented an an angle $\theta_{AB}$ to Bob's, (\ref{singlet}) can be written
\be
\label{singlet2}
|\psi^{\mathrm{singlet}}_{AB}\rangle=\cos\frac{\theta_{AB}}{2} \left ( \frac{|\ 1_A\rangle |-1_B\rangle- |-1_A\rangle |\ 1_B\rangle}{\sqrt 2} \right )+ \sin\frac{\theta_{AB}}{2} \left ( \frac{|\ 1_A\rangle |\ 1_B\rangle - |-1_A\rangle |-1_B\rangle}{\sqrt 2} \right ) 
\ee
Providing $\theta_{AB}$ satisfies (\ref{rat}), the singlet state (\ref{singlet2}) in RaQM is represented, for a specific run of a Bell experiment, by the 2 length-$L$ bit strings
\be
\label{bitstrings}
|\psi^{\mathrm{singlet}}_{AB}\rangle_\xi=\left\{
\begin{array} {c}
\xi \{\underbrace{\ 1,\; \ \; 1,\ \; 1,\ \ \dots \ \ 1,\ \; 1,\ \; 1,\ \ -1,-1,-1, \ \ldots \ ,-1,-1,-1}_{\frac{\pi}{2}, 0}\} \\
\xi \{\underbrace{1,1,\; \ldots , 1, \ -1, -1,\ldots,-1,}_{\pi-\theta_{AB}, 0} \ \underbrace{\ 1,1, \ldots ,\ \ 1,\ -1,-1 \ldots \ ,-1}_{\theta_{AB}, 0}\} 
\end{array}
\right.
\ee
The measurement outcomes are given by some $M(\xi)$th pair of bits prior to the application of $\xi$. As above, we can assume that $M(\xi)$, for all practical purposes random, is fixed at the time the entangled particles leave their source region, and hence is `known' to both qubits. In particular, when $\theta_{AB}=0$, then if Alice measures $+1$, Bob will certainly measure $-1$ and \emph{vice versa}. 

Following EPR \cite{EPR} and Bell \cite{Bell:1964}, we will say that a theory is EPR/Bell nonlocal if, for a particular run of a Bell experiment, Alice's measurement outcome can be affected by Bob's measurement setting, and \emph{vice versa}. That is to say, if the theory predicts that, were Bob were to change his mind and measure differently and this change of mind affects Alice's measurement outcome, then, since Alice and Bob can be arbitrarily far apart, the theory is nonlocal. By contrast, if the theory predicts that Alice's measurement outcome is unaffected by Bob's change of measurement setting, then the theory is local. 

It is easy to see that RaQM is EPR/Bell local. Let us suppose, for a particular entangled particle pair, Alice chooses measurement orientation $\boldsymbol A_{\mathrm{nom}}$ and Bob $\boldsymbol B_{\mathrm{nom}}$, where the exact settings satisfy $\boldsymbol A \cdot \boldsymbol B$ satisfies (\ref{rat}). If Bob is a free agent, then he could have chosen any other direction $\boldsymbol B'_{\mathrm{nom}}$ providing there exist an exact setting $\boldsymbol B'$ where $\boldsymbol A \cdot \boldsymbol B'$ satisfies (\ref{rat}). In this way, from (\ref{bitstrings}), if Bob counterfactually changes his measurement orientation from $\boldsymbol B_{\mathrm{nom}}$ to $\boldsymbol B'_{\mathrm{nom}}$, Alice's bit string does not change. In particular, the $M(\xi)$th bit, corresponding her measurement outcome, does not change. 

Since both bit strings in (\ref{bitstrings}) have equal numbers of 1s and -1s, (\ref{bitstrings}) is equivalent to 
\be
\label{bitprime}
|\psi_{AB}\rangle_\xi'=\left\{
\begin{array} {c}
\xi' \{\underbrace{1,1,\; \ldots , 1, \ -1, -1,\ldots,-1}_{\pi-\theta_{AB}, 0} \ \underbrace{\ 1,1, \ldots ,\ \ 1,\ -1,-1 \ldots \ ,-1}_{\theta_{AB}, 0}\} \\
\xi' \{\underbrace{\ 1,\; \ \; 1,\ \; 1,\ \ \dots \ \ 1,\ \; 1,\ \; 1,\ \ -1,-1,-1, \ \ldots \ ,-1,-1,-1}_{\frac{\pi}{2}, 0}\} 
\end{array}
\right.
\ee
where $\xi'$ is such that the two representations (\ref{bitstrings}) and (\ref{bitprime}) are identical. Hence, the measurement outcomes correspond to the $M(\xi)$th pair of bits in (\ref{bitstrings}), equal to the $M(\xi')$th pair of bits in (\ref{bitprime}), where $0 \le M \le L$. The form (\ref{bitprime}) allows Alice\'s measurement settings to vary from $\boldsymbol A_{\mathrm{nom}} \mapsto \boldsymbol A'_{\mathrm{nom}}$. Providing the corresponding exact settings satisfy $\boldsymbol A' \cdot \boldsymbol B \in \mathbb Q$, she can do this without affecting Bob's measurement outcome. 

The rationality restrictions on $\boldsymbol A \cdot \boldsymbol B$, $\boldsymbol A \cdot \boldsymbol B'$ and $\boldsymbol A' \cdot \boldsymbol B$ do not impose any constraint on the ability of experimenters to freely choose their measurement settings. When Alice chooses some nominal direction $\boldsymbol A_{\mathrm{nom}}$, and Bob some nominal direction $\boldsymbol B_{\mathrm{nom}}$, their choice defines, not a pair of single points on the celestial sphere, but a pair of small neighbourhoods of points. Within these neighbourhoods there will certainly will exist exact measurement orientations $\boldsymbol A$ and $\boldsymbol B$ such that $\boldsymbol A \cdot \boldsymbol B \in \mathbb Q$. If Bob counterfactually changes his mind, $\boldsymbol B'_{\mathrm{nom}}$ again denotes a small neighbourhood of points on the celestial sphere, and again there will exist pairs of points, one from $\boldsymbol A$ and one from $\boldsymbol B'$, which satisfy the rationality condition $\boldsymbol A \cdot \boldsymbol B' \in \mathbb Q$. In this way, for large enough $L$, the rationality conditions (\ref{rat}) do not impose any restriction on Bob changing his mind, keeping Alice's choice fixed, and \emph{vice versa}. From this point of view, experimenters can freely choose their measurement settings and the effects of this change of mind are only locally felt. 

The key question to be asked is whether such a local theory satisfies Bell's inequality. In Section \ref{Bell} we show that the rationality constraints on the exact measurement settings prevent Bell's inequality from being satisfied. The argument hinges around an extension of the discussion above, where we are required to consider not one but two simultaneous counterfactuals. 

However, because the argument (based as it is on number theory) may seem unusual and perhaps even esoteric in a quantum physics setting, as a warm up to Bell's Theorem we first discuss quantum properties of single particles from RaQM's perspective. The number-theoretic argument used to show how Bell's inequality is violated is, at heart, \emph{identical} to explaining complementarity and non-commutativity of observables in single-particle RaQM - supporting Feynman's claim that there is only one real mystery in quantum physics (in this case, a mystery that can be unravelled by invoking number-theoretic properties of trigonometric functions). If it can be accepted that number-theoretic incommensurateness can explain single-particle non-commutativity, then we may be easier to accept that it can also explain the violation of Bell's inequality. 

\section{Complementarity and Non-commutativity from Number Theory}
\label{complementarity}

As a warm-up to Bell's Theorem, we analyse the quantum nature of single-particle systems in RaQM. 

\subsection{Complementarity}

\begin{figure}
\centering
\includegraphics[scale=0.5]{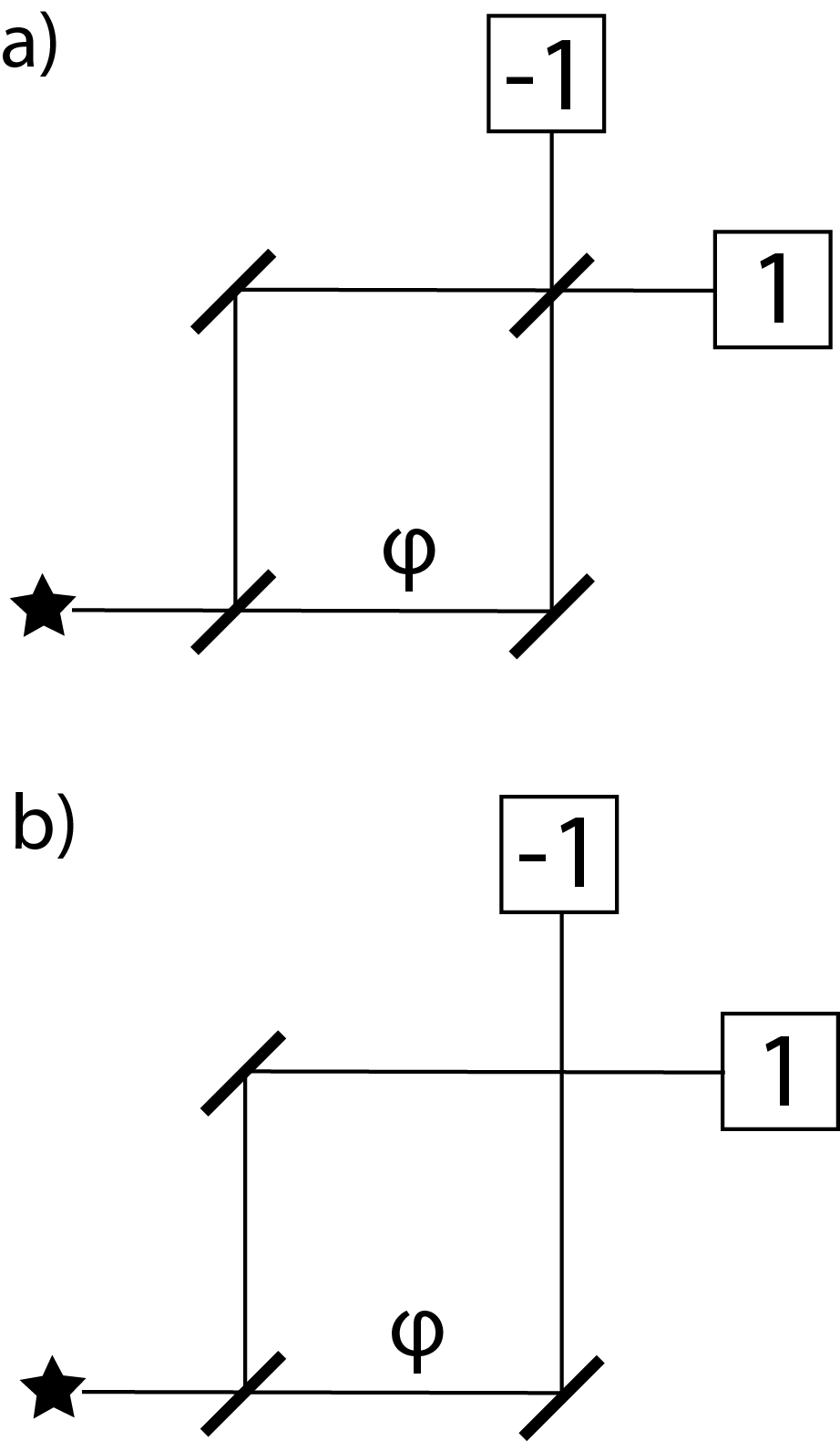}
\caption{\emph{In QM, complementarity implies that simultaneous measurement outcomes do not exist for a quantum system in a Mach-Zehnder interferometer with a) the second half silvered mirror in place (a wave-like interferometric measurement), and b) removed (a particle-like which-way measurement). In RaQM, such complementarity is a mathematical consequence of a number-theoretic incommensurateness between angles and their cosines: Niven's Theorem \cite{Niven}.}}
\label{MZ}
\end{figure}

Consider a Mach-Zehnder interferometer. In QM, the qubit state of a photon in the interferometer's measurement basis is
\begin{equation}
\label{MZ1}
\cos\frac{\phi}{2} |1\rangle + \sin\frac{\phi}{2} |-1\rangle
\end{equation}
where $\phi$ denotes a phase difference between the two arms of the interferometer (Fig \ref{MZ}a). By contrast, if the interferometer's second half-silvered mirror is removed (Fig \ref{MZ}b), the qubit state of the photon in the measurement basis is
\begin{equation}
\label{MZ2}
\frac{1}{\sqrt 2}(|1\rangle+ e^{i \phi} |-1\rangle)
\end{equation}
in the measurement basis, and a `which-way' measurement is made. 

According to the discretisation (\ref{rat}), if an experimenter $E$ freely chooses to perform an interferometric measurement, then necessarily $\cos^2 (\phi/2)$ in (\ref{MZ1}), and hence $\cos \phi$, is a rational number. However, for the photon state to be well defined in the counterfactual measurement basis where $E$ removes the half-silvered, from (\ref{MZ2}) and (\ref{rat}) we additionally require that $\phi/\pi$ is rational. 

However, we can now invoke Niven's Theorem \cite{Niven} which states that the only values $0 \le \phi < 2\pi$ where $\cos \phi$ and $\phi/\pi$ are simultaneously rational are:
\be
 \phi =0, \frac{\pi}{3},  \frac{\pi}{2},  \frac{2\pi}{3},  \pi, \frac{4\pi}{3},  \frac{3\pi}{2},  \frac{5\pi}{3}.
 \ee
Suppose $E$ freely chooses $\phi_{\mathrm{nom}}=0$, i.e. to experimental tolerance the arms of the interferometer have equal length. Of course, $E$ cannot guarantee that the exact phase difference is zero - a gravitational wave from some distant galaxy may pass through the interferometer as the experiment is being performed. Now, according to our model of discretised Hilbert Space, there are $L$ discrete phases around the circle. If $L$ is big enough (according to \cite{Palmer:2025a}, $L \approx 10^{100}$ for a typical qubit in a quantum computer), there will be $O(L)$ discrete phases in any nominal neighbourhood of $\phi=0$. For only one of the exact values in such a neighbourhood ($\phi=0$ precisely) is $\cos \phi$ and $\phi/\pi$ simultaneously rational. Hence, with overwhelming likelihood, the quantum state in the counterfactual world where the basis is measuring particle-like properties of the photon, cannot be simultaneously well-defined with the real-world basis where an interferometric measurement was performed, and \emph{vice versa}. That is to say, in our deterministic model with rational number constraints (\ref{rat}), complementarity is implied by simple number-theoretic incommensurateness (something completely concealed by the continuum nature of state space in QM). 

\subsection{Non-Commutativity}

\begin{figure}
\centering
\includegraphics[scale=0.7]{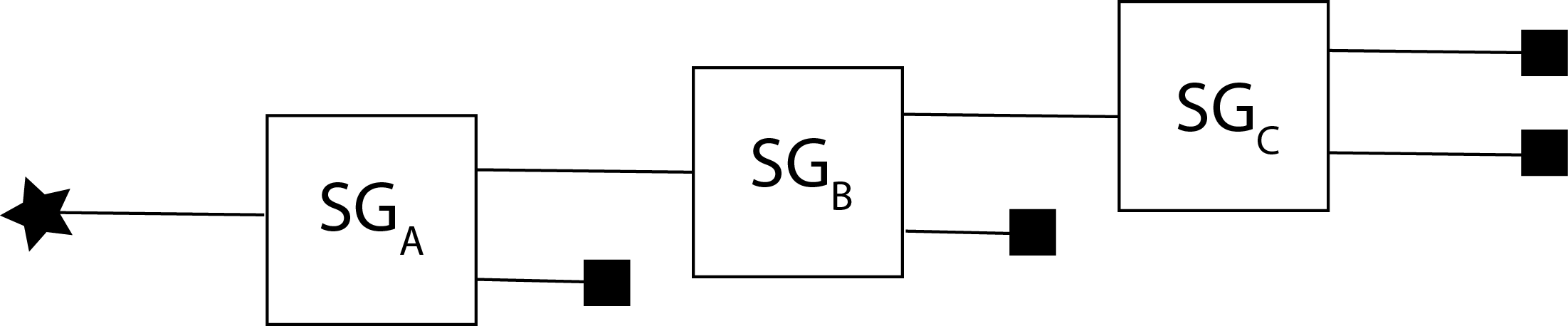}
\caption{\emph{A spin-$1/2$ quantum system is prepared `up' by Stern-Gerlach device $SG_A$. The spin-up output of $SG_A$ is fed into device $SG_B$ and the spin-up output of $SG_B$ is fed into $SG_C$. The experimenter is free to choose the nominal orientations of these apparatuses as they like. However, according to the rationality constraints (\ref{ratSG2}), Niven's Theorem prevents the simultaneous counterfactual world where $SG_B$ and $SG_C$ are swapped from having a well-defined measurement outcome.}}
\label{SG}
\end{figure}

\begin{figure}
\centering
\includegraphics[scale=0.5]{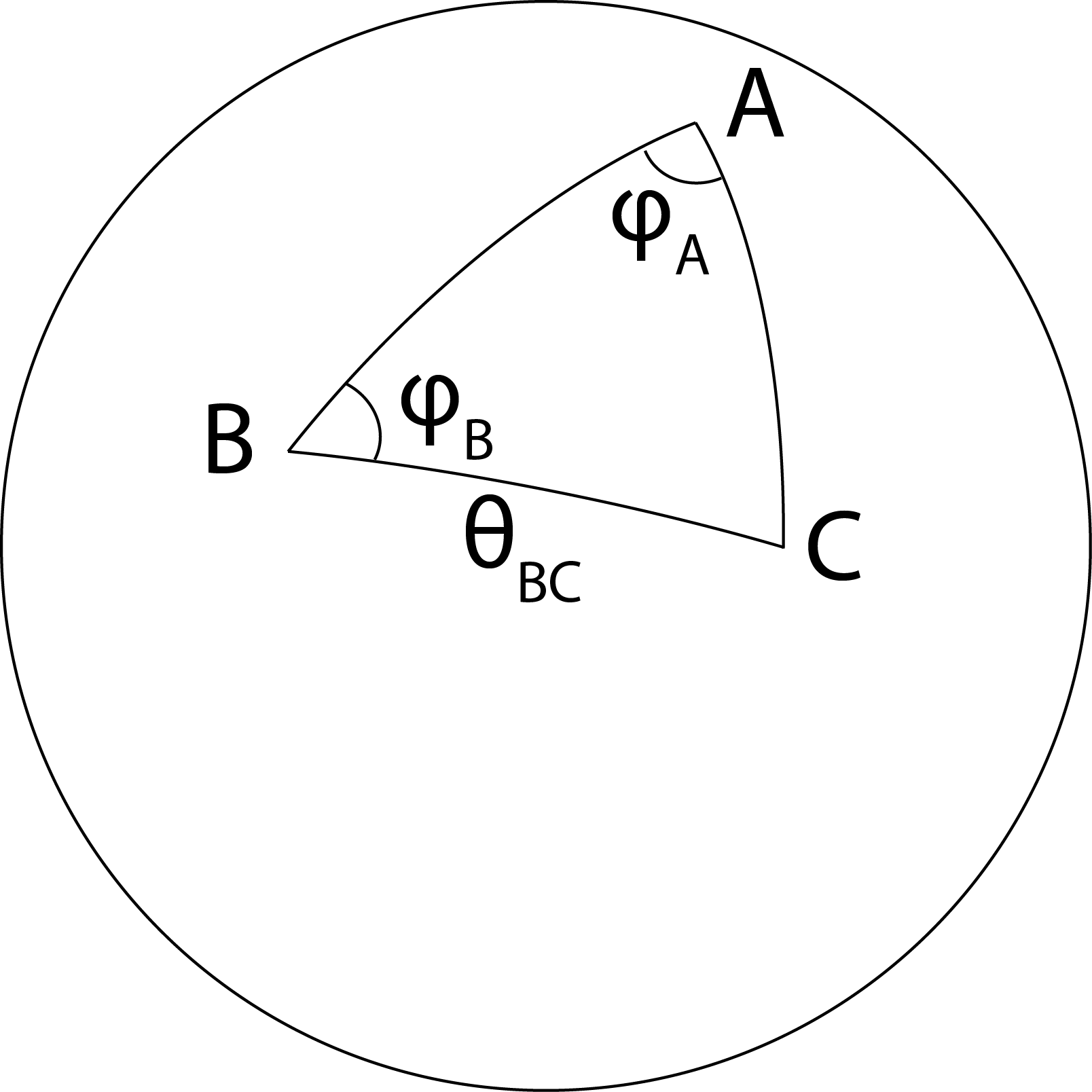}
\caption{\emph{An impossible spherical triangle, where the cosine of the angular length of each side of the triangle on the unit sphere is rational, and the internal angles are rational in degrees. The impossible triangle provides the basis for understanding both single particle non-commutativity and violation of Bell's inequality for entangled particles. In practice, for the sequential SG experiment, $\phi_B$ is close but not exactly equal to $180^\circ$.}}
\label{triangle}
\end{figure}

By passing a spin-1/2 particle through a Stern-Gerlach ($SG_A$) device oriented along $\boldsymbol A$, a particle is prepared with spin up relative to some direction $\boldsymbol A$,  (see Fig \ref{SG}). The particle is then passed through a second Stern-Gerlach device ($SG_B$) along direction $\boldsymbol B$. The spin-up output of $SG_B$ is then passed to a third Stern-Gerlach device ($SG_C$) oriented along direction $\boldsymbol C$. Again, $E$ is free willed and orientates the SG devices (to some nominal accuracy) as they like. Let $A$, $B$ and $C$ denote vertices of a spherical triangle $\triangle ABC$ on the unit (celestial) sphere, corresponding to the tips of the three unit vectors $\boldsymbol A$, $\boldsymbol B$ and $\boldsymbol C$ in physical space. See Fig \ref{triangle}. 

We will assume that a given quantum system (`particle') whose spin is prepared `up' relative to $SG_A$, has spin $S(\lambda, \boldsymbol A, \boldsymbol B)$ coming out of $SG_B$ where $S$ is some deterministic function returning either $+1$ or $-1$ and $\lambda$ is a hidden variable. From the discussion in Section \ref{discrete}, $\lambda$ is related to $\xi$. Similarly, the same quantum system prepared `up' relative to $SG_B$, has spin $S(\lambda, \boldsymbol B, \boldsymbol C)$ coming out of $SG_C$. By definitiion, $\boldsymbol A \cdot \boldsymbol B = \cos \theta_{AB}$ where $\theta_{AB}$ is the angular distance between $A$ and $B$ on the unit sphere, etc.

Although $\boldsymbol A$, $\boldsymbol B$ and $\boldsymbol C$ will be approximately coplanar, they will not be exactly coplanar. Hence $\triangle ABC$ is non-degenerate, but where $\phi_B \approx \pi$.  Following the discussion in Section \ref{discrete}, the rationality constraints (\ref{rat}) imply that 
\be
\label{ratSG2}
\cos \theta_{AB} \in \mathbb Q;\ \ 
\cos \theta_{BC} \in \mathbb Q.
\ee
In addition, the angle $\phi_B$ between the great circles AB and BC must satisfy
\be
\label{ratSG3}
\frac{\phi_B}{2\pi} \in \mathbb Q.
\ee
(Hence the points A and C both lie on a discretised Bloch Sphere with pole at B.) The non-commutativity of spin observables in QM can now be framed as follows: what prevents $SG_B$ and $SG_C$ being counterfactually swapped, keeping the quantum system fixed? That is to say, although in reality the particle was sent from $SG_A$ to $SG_B$ and then to $SG_C$, what's to prevent a counterfactual world where everything is the same, except that the quantum system is sent from $SG_A$ to $SG_C$ and then to $SG_B$? For this counterfactual world to be simultaneously well-defined, we require, in addition to (\ref{ratSG2}) and (\ref{ratSG3}),
\be
\boldsymbol A \cdot \boldsymbol C \equiv \cos \theta_{AC} \in \mathbb Q
\ee
Hence our question can be framed as follows: what's to stop the cosines of the angular lengths of all three sides of the spherical triangle $\triangle ABC$ in Fig \ref{triangle} from being rational numbers, if, in addition, at least one of the internal angles is also rational (in degrees)? 

By way of contradiction, let's suppose this is possible. The cosine rule applied to the triangle $\triangle ABC$ can be written
\be
\cos \theta_{AC}= \cos \theta_{AB} \cos \theta_{BC} + \sin \theta_{AB} \sin \theta_{BC} \cos \phi_B
\ee
If the left hand side of this equation and the first term on the right hand side are both rational, then the second term on the right hand side must be rational. Squaring, this implies that 
\be
(1-\cos^2 \theta_{AB})(1- \cos^2 \theta_{BC}) \cos^2 \phi_B
\ee
must be rational, and therefore $\cos^2 \phi_B$ and hence $\cos 2\phi$ must be rational. But we can again appeal to Niven's Theorem to derive a contradiction. If the three SG devices are approximately coplanar (i.e. $\phi_B \approx \pi$ to nominal accuracy), then there is only one exact angle ($\phi_B=\pi$) in any nominal neighbourhood of $\phi_B=\pi$ where $\cos 2\phi_B$ and $\phi_B/2\pi$ are simultaneously rational. However, in the neighbourhood of $\phi_B=\pi$, there are $O(L)$ possible values in each of which $\phi_B/2\pi$ is rational but $\cos 2\phi_B$ is irrational. As above, we estimate $L \approx 10^{100}$ \cite{Palmer:2025a}.  

Hence, with absolutely overwhelming probability, we have our contradiction: the angular lengths of the three sides of the triangle can't all have rational cosines if the internal angles are rational in degrees.  Hence the counterfactual world where $SG_B$ and $SG_C$ are swapped, keeping everything else fixed, cannot be well defined, simultaneous with the real world, even though $E$ was free to choose measurement orientations as they liked. That is to say, in our deterministic model with rational number constraints, the non-commutativity of observables is implied by Niven's Theorem.

\section{$p$-adic Numbers and Fractal Geometry}
\label{2adic}

The discussion above may provoke an immediate objection: since the irrational reals are infinitesimally close to the rationals, any restriction to the rationals must surely imply an exceptionally finely tuned theory. Physical theories should be robust to small perturbations. From this perspective, RaQM does not seem robust. That this is not the case raises an important issue. From a mathematical point of view, irrational real numbers are defined as the limits of Cauchy sequences of rational numbers with respect to the Euclidean metric. However, there is another (indeed only one other) class of metric on the rationals, the $p$-adic metric \cite{Katok}; with respect to this metric, irrational reals are simply undefined, like quantum states in irrational bases in this paper. The $p$ adic numbers are the limits of Cauchy sequences of rationals with respect to the $p$-adic metric. From a geometric point of view, the $p$-adic metric is to fractal geometry as the Euclidean metric to Euclidean geometry. In particular, the set of 2-adic numbers is homeomorphic to the simple Cantor Set \cite{Katok}. (This is why I got excited when I saw the picture of a fractal on the front cover of \cite{HileyPeat}.) As an example, putting $-1 \mapsto 0$ we can write (\ref{bitstrings}) with $\theta_{AB}=\pi/2$ as the pair of 2-adic numbers 
\be
\label{bitstring5}
|\psi^{\mathrm{singlet}}_{AB}\rangle_\xi=\left\{
\begin{array} {c}
\xi(1111 \ldots 11110000 \ldots 0000). \\
\xi (1111 \ldots 00001111 \ldots 0000).
\end{array}
\right.
\ee
or as the 4-adic number $= \xi(3333 \ldots 22221111 \ldots 0000)$. 

In \cite{Palmer:2025a}, the process of measurement is represented as a reduction in the information content of the quantum state by $L-1$ bits, at a rate of 1 bit per Planck time, to the classical limit where $L=1$. This can be represented by the application of a chaotic shift map  \cite{WoodcockSmart} (dividing the 2-adic numbers by 2 at each step, and disregarding the fractional part). For example, with $L=8$, suppose after the application of $\xi$ we have $|\psi\rangle =.10011010.$. Then under state reduction
\be
|\psi\rangle= 10011010. \mapsto 1001101.  \mapsto 100110. \mapsto 10011.
 \mapsto 1001. \mapsto 100. \mapsto 10. \mapsto 1.
\ee
in 7 Planck times. In this way, state vector collapse can also be viewed geometrically in terms of a fractal zoom, or equivalently as a chaotically evolution to one of two (or more) stable attractors \cite{Palmer:2025a} \cite{Palmer:1995}. With $L \approx 10^{100}$, the collapse time is longer than the age of the universe. In \cite{Palmer:2025a} the dependence of $L$ on the mass/energy of the qubit is discussed, providing shorter collapse times for more massive qubits. 

In addition, as a quantum system progressively entangles (decoheres) with the environment, the total quantum state becomes representable as a $p$-adic number where $p \ggg 1$. In the context of $p$-adic number theory, the reals are sometimes represented as numbers at $p=\infty$. In this context, the reduction to $L=1$ corresponds to evolution to a single real number - the way a classical system is represented. 

The key point here is that whilst the relevant metric to define distances between objects in physical space is the Euclidean metric, the relevant metric to define distances between different worlds in state space (e.g. one real, the other counterfactual) is the $p$-adic metric. 

\section{Bell's Theorem}
\label{Bell}

Finally, we can analyse Bell's inequality \cite{Bell:1964} 
\be
\label{bellineq}
1 \ge
|Co(\boldsymbol A_{\mathrm{nom}}, \boldsymbol B_{\mathrm{nom}}) -Co(\boldsymbol A_{\mathrm{nom}}, \boldsymbol C_{\mathrm{nom}})| -Co(\boldsymbol B_{\mathrm{nom}},\boldsymbol C_{\mathrm{nom}})
\ee
using \emph{exactly} the same `impossible triangle' argument we used to analyse single-particle non-commutativity in RaQM. Analysing the CHSH inequality is fundamentally no different \cite{Palmer:2024} though a little more complicated. In (\ref{bellineq}), $Co$ denotes a correlation over many experimental runs of individual quantum systems, each prepared in the same singlet state (\ref{singlet}), each run associated with a specific $\lambda$.  

Experimentally, the correlations in (\ref{bellineq}) are determined from measurements on three separate sub-ensembles of particles. Since, as above, experimenters can only set measuring orientations to nominal accuracy, the exact settings $\boldsymbol A$, $\boldsymbol B$ and $\boldsymbol C$ will not be the same for each run of the Bell experiment. For the $i$th run where the first correlation in (\ref{bellineq}) is being experimentally found, RaQM requires that the exact settings satisfy:
\be
\label{e1}
\boldsymbol A_i \cdot \boldsymbol B_i \equiv \cos \theta_{A_iB_i} \in \mathbb Q
\ee
Similarly, for the $i$th run where the second correlation of (\ref{bellineq}) is being experimentally found:
\be
\label{e2}
\boldsymbol A_i \cdot \boldsymbol C_i \equiv \cos \theta_{A_iC_i} \in \mathbb Q
\ee
Finally, for the $i$th run where the third correlation of (\ref{bellineq}) is being experimentally found:
\be
\label{e3}
\boldsymbol B_i \cdot \boldsymbol C_i \equiv \cos \theta_{B_iC_i} \in \mathbb Q.
\ee
Since (\ref{e1}), (\ref{e2}) and (\ref{e3}) correspond to separate ensembles of particles, these rationality conditions can be trivially satisfied. With this, RaQM predicts
\begin{align}
Co(\boldsymbol A_{\mathrm{nom}}, \boldsymbol B_{\mathrm{nom}})&= -\boldsymbol A_{\mathrm{nom}} \cdot \boldsymbol B_{\mathrm{nom}},  \nonumber \\
Co(\boldsymbol A_{\mathrm{nom}}, \boldsymbol C_{\mathrm{nom}})&= -\boldsymbol A_{\mathrm{nom}} \cdot \boldsymbol C_{\mathrm{nom}}, \nonumber \\
Co(\boldsymbol B_{\mathrm{nom}}, \boldsymbol C_{\mathrm{nom}})&= -\boldsymbol B_{\mathrm{nom}} \cdot \boldsymbol C_{\mathrm{nom}}
\end{align}
 exactly as in QM, and Bell's inequality can be violated for suitable choices of nominal measurement setting. 
 
Since RaQM is locally real, what aspect of RaQM negates the proof that RaQM must satisfy Bell's inequality? For a putative local hidden-variable model to satisfy Bell inequalities, we integrate (or sum)
\be
\label{bellsum}
|S(\lambda, \boldsymbol A)S(\lambda, \boldsymbol B)- S(\lambda, \boldsymbol A)S(\lambda, \boldsymbol C)|-S(\lambda, \boldsymbol B)S(\lambda, \boldsymbol C)
\ee
over $\lambda$, for the associated local deterministic spin functions $S(\lambda, \boldsymbol A)$ etc. But it is only possible to perform this sum if all terms in (\ref{bellsum}) are mathematically well defined, for each $\lambda$. 

We now refer to the same spherical triangle as in Fig \ref{triangle}. For a given run (i.e. fixed $\lambda$) of a Bell experiment suppose Alice chose $\boldsymbol A_{\mathrm{nom}}$ and Bob $\boldsymbol B_{\mathrm{nom}}$, with corresponding exact settings $\boldsymbol A$ and $\boldsymbol B$. Then, in order for (\ref{bellsum}) to be well defined, it must be the case that the two following counterfactual worlds \emph{simultaneously} have well-defined measurement outcomes with the real world: one where Alice's exact setting continues to be $\boldsymbol A$ and Bob chooses $\boldsymbol C_{\mathrm{nom}}$ with some exact setting $\boldsymbol C$, and one where Bob's exact setting continues to be $\boldsymbol C$ and Alice's exact setting is $\boldsymbol B$. Hence, in order for (\ref{bellsum}) to be well defined, we require that, simultaneously:
\be
\label{ratBell}
\cos \theta_{AB} \in \mathbb Q;\ \ 
\cos \theta_{AC} \in \mathbb Q;\ \ 
\cos \theta_{BC} \in \mathbb Q
\ee
Since $C$ is arbitrary within the neighbourhood defined by $\boldsymbol C_{\mathrm{nom}}$,  the first counterfactual can be satisfied trivially. 

However, crucially, the rationality conditions cannot now be satisfied for the second counterfactual. To see this, note that although $\boldsymbol  A_{\mathrm{nom}}$, $\boldsymbol  B_{\mathrm{nom}}$ and $\boldsymbol  C_{\mathrm{nom}}$ are coplanar, the corresponding exact angles $\phi_A$, $\phi_B$ and $\phi_C$ will not be \emph{precisely} equal to $0^\circ$ or $180^\circ$. In addition to (\ref{ratBell}), we must additionally demand that $\phi_A$ (only approximately $0^\circ$ or $180^\circ$) is rational in degrees. But now the impossible triangle corollary can again be invoked to imply that it is impossible for the quantum state, not only in the real-world basis, but also in \emph{both} of counterfactual-world bases to be simultaneously well defined. Of course this conclusion applies no matter what the real world nominal measurement directions are (i.e., they could be $\boldsymbol  A_{\mathrm{nom}}$ and $\boldsymbol  C_{\mathrm{nom}}$, or $\boldsymbol  B_{\mathrm{nom}}$ and $\boldsymbol  C_{\mathrm{nom}}$). Hence, (\ref{bellsum}) is generally undefined. As discussed in Section \ref{discrete}, neither realism nor locality are violated here. Rather, the MI assumption is violated, not for the freely-chosen nominal settings, but for the exact settings that were never under the experimenters' control anyway. 

We have described RaQM in terms of hidden variables. Does this mean that RaQM is somehow classical? There are (at least) three reasons why not:
\begin{itemize}
\item A classical system is computational in nature. However, with $L = 10^{100}$ it would be impossible to determine whether a putative state satisfies the rationality conditions (\ref{rat}) within the age of the universe (even when one step of an algorithm is performed in a Planck time). Moreover, it is impossible to determine whether (\ref{rat}) is satisfied by a simpler algorithm that \emph{can} be executed within the age of the universe. 
\item Consistent with $L=10^{100}$, an experiment performed on a qubit always returns an outcome $\pm 1$. The `irrational' directions where states and hence outcomes are undefined are too small scale in physical space to be detectable. Just as it is impossible to construct a measurement device which has the resolution to detect these `undefined' irrational directions, it would be impossible to manufacture a classical system with these rational constraints properties. It is worth noting that the vast magnitude of $L$ is intrinsically linked with the weakness of gravity. The world may ultimately be a quantum world and not a classical world because gravity is so weak. 
\item As discussed, a deterministic dynamical system describing RaQM is $p$-adic in nature. Classical dynamical systems, by contrast, are based on real numbers and the associated Euclidean metric. 
\end{itemize}
In the classical $L=1$ limit of RaQM, experiment settings are entirely passive and merely reveal pre-existing spin orientations for the system being measured (the holistic nature of the laws of physics break down). As such, there are no restrictions on counterfactual measurement orientations and the arguments here (for the violation of Bell's inequality) fail in the classical limit (where $G=0$ precisely). 

\section{Discussion and Conclusions}
\label{conc}

RaQM is a nascent (and still developing) theory of quantum physics based on the notion that the state space of quantum physics is not the continuum Hilbert Space of QM, but is a gravitationally induced discretisation of Hilbert Space, for which the continuum is a singular limit at $G=0$. In this theory, the Schr\"{o}dinger equation is not itself modified. However, the quantum state is a basis-independent construct, only defined in bases which satisfy certain rational-number constraints. 

Using RaQM, we have described a novel interpretation of Bell's Theorem as a violation of Measurement Independence, not for nominal measurement settings under the experimenters' full control, but for exact measurement settings which were never under their control anyway, not least due to the (unshieldable) effects of gravitational waves from distant sources. This particular interpretation of Bell's Theorem is reminiscent of the role of finite precision measurements in relation to the Kochen-Specker Theorem \cite{Meyer:1999}.

According to RaQM, the real meaning behind Bell's Theorem is not that the laws of physics are EPR/Bell nonlocal (i.e., that measurement outcomes here can depend on measurement settings there, seemingly implying some kind of infinite-speed quantum communication), or other weird ideas (conspiracy, infinite-speed communication, wormholes, anti-realism, retrocausality, branching of worlds etc), but instead that such laws are holistic.  An example of such holism is Mach's principle: that inertia here is due to mass there. Of course, as all physicists fully understand, there is nothing nonlocal about Mach's principle; indeed it was a key motivation for Einstein in formulating his locally causal theory of general relativity. The author has proposed the Invariant Set Postulate \cite{Palmer:2009a} as an example of a holistic geometric law of quantum physics, that the universe is a deterministic chaotic system evolving precisely on some fractal invariant set in state space. This postulate is consistent with the gappy Hilbert-Space structure proposed here, and with the chaotic shift map describing state reduction and measurement. If this postulate is correct, it means that we should not think of the large-scale structure of the universe as emergent from laws which operate only on the small scale (the philosophy underpinning the Standard Model).  Instead, the laws of physics (defined geometrically in state space) directly determine both the large-scale and small-scale structure of space-time. 

Put like this, building bigger and bigger particle colliders, in the hope that this will lead us to a synthesis of quantum and gravitational physics, and ulimately a Theory of Everything, may be a fruitless exercise. A more productive route to uncovering such a theory may be trying to discover experimentally the limits of QM itself \cite{Palmer:2025a}, and the possible role of gravity in the measurement process. 

This notion of holism is very much embodied in the title of Bohm and Hiley's influential text book: The Undivided Universe \cite{BohmHiley}. The book is devoted to an exposition of the de Broglie-Bohm interpretation of QM. However, as is well known, de Broglie-Bohm theory is explicitly nonlocal (and for that reason is not well accepted in the physics community). This raises the question: how would one modify de Broglie-Bohm theory for it to be consistent with RaQM and its discretised Hilbert Space?

The answer lies in modifying the structure of the quantum potential, a square-integrable function on configuration space in de Broglie-Bohm theory. Precisely because it is square integrable, the quantum potential does not have the fine-scale `gappy' state-space property of RaQM, and de Broglie-Bohm is consequently EPR/Bell nonlocal. However, suppose the de Broglie-Bohm quantum potential is merely a smoothed coarse-grained approximation for some deeper, gappy, fractal state-space geometry. Then, in the `exact' fine-grained theory, one may be able to bring to bear the arguments in this paper to show that modified de Broglie-Bohm theory is not EPR/Bell nonlocal after all, though still with all the attractive features of Bohm and Hiley's `Undivided Universe'. The key to accepting such a possibility (in contrast with the beliefs of Many-Worlds advocates) is to eschew the idea that the continuum theory we call QM is the final word in quantum physics. 

The idea of modifying the quantum potential to allow it  to describe fine-scale geometric structure was the basis of my email to Basil all those years ago. The idea certainly appealed to him, and this in turn motivated me to make my arguments more rigorous and watertight, which I (finally) believe them to be today. In Basil's last email communication with me, he expressed his excitement to have discovered that de Witt \cite{DeWitt:1952}, years earlier, had in fact given a geometric description of the quantum potential. Not only that, but DeWitt's paper appeared in the same volume as Bohm landmark paper on hidden variables and the de Broglie-Bohm interpretation \cite{Bohm:1952}! We were finally starting to converge on a geometric interpretation of de Broglie-Bohm theory. However, De Witt's geometric representation was Riemannian and not $p$-adic. So the final convergence, if there is to be one, lies in the future. I am optimistic we will get there, but some of the details still elude me. 

Thank you, Basil, for years of stimulating interaction with you. 

\section*{Acknowledgments}

I would like to thank Georg Wikman for his extraordinary hospitality in hosting some of the most outstanding summer meetings on quantum foundations in his summer house in Sweden (during which I got to know Basil well, and where I met some of the leaders in the field). Preparing talks for these meetings was an enormous stimulus to thinking hard about the problems at hand. My thanks also to Michael Wright, not only for his many insightful comments during these workshops, not least about the all-important interface between physics and mathematics, but who recorded all of these meetings. His project to put his many recordings online is something I look forward to seeing completed, hopefully in the not too distant future. For giving me the confidence to question quantum mechanics, I acknowledge the inspirational work of Roger Penrose. Roger and my PhD supervisor Dennis Sciama were my two key mentors during my PhD years at Oxford. In later years I was also inspired by the MIT meteorologist Ed Lorenz, whose discovery of the fractal geometry of chaos was one of the greatest in science, in my view. I would like also to thank \v{C}aslav Brukner and Luis C. Barbado for recent insightful discussions and correspondence, and anonymous reviewers for helpful comments on a first draft of this paper. 

\bibliography{mybibliography}

\section*{Appendix}
\label{secA1}
\subsection*{Complex Numbers and the Discretised Bloch Sphere}
Complex numbers play a vital role in QM. On the discretised Bloch Sphere, discretised complex numbers denote permutation/negation operators acting on length $L$ bit strings $\{a_1, a_2, \ldots a_L\}$, where $a_i \in \{1,-1\}$ and
\begin{equation}
-\{a_1, a_2, \ldots a_L\}=\{-a_1, -a_2, \ldots -a_L\}
 \end{equation}
 We start with the $2 \times 2$ matrix
\begin{equation}
\label{I}
i =
\left( \begin{array} {cc}
0& 1 \\
- 1 &0
\end{array} \right),
\end{equation}
whence
 \begin{align}
 i \{a_1, a_2\}^T&=\{a_2, -a_1\}^T \nonumber \\
 i^2 \{a_1, a_2\}^T&=-\{a_1, a_2\}^T 
 \end{align}
 With $4 \mid L$ we construct the three $L \times L$ matrices
 \begin{equation}
\label{I}
J_1 =
\left( \begin{array} {cc | cc}
\ I&\ &\ &\ \\
\ &\ I&\ &\ \\
\hline
\ &\ &-I&\ \\
\ &\ &\ &-I 
\end{array} \right);\ \ 
J_2 =
\left( \begin{array} {cc | cc}
\ &\ &\ 1&\ \\
\ &\ &\ &-1\\
\hline
-1&\ &\ &\ \\
\ &\ 1&\ &\  
\end{array} \right);\ \ 
J_3 =
\left( \begin{array} {cc | cc}
\ &\ &\ I&\ \\
\ &\ &\ &-I\\
\hline
\ I&\ &\ &\ \\
\ &-I&\ &\  
\end{array} \right)
\end{equation}
where $I$ is the $L/4 \times L/4$ matrix
 \begin{equation}
\label{I}
I =
\left( \begin{array} {cccc}
\ i&\ &\ &\ \\
\ &\ i&\ &\ \\
\ &\ &\ddots&\ \\
\ &\ &\ &i 
\end{array} \right). 
\end{equation}
where $1$ is the $L/4 \times L/4$ unit matrix. It is readily shown that $\{J_1, J_2, J_3\}$ satisfy the quaternionic relations
\begin{equation}
J_1^2=J_2^2=J_3^2=- 1_L; \ \ \  J_1 \times J_2=J_3.
\end{equation}
and $1_L$ is the $L \times L$ unit matrix. 

With this, the discretised single qubit state (\ref{qubit3}) can be written
\begin{equation}
\label{state}
|\psi(\theta_m,\phi_n) \rangle \equiv \zeta^n [\mathscr I_L(m)] \bmod \xi
\end{equation}
where 
\begin{equation}
\label{Lbits}
\mathscr I_L(m)=\{\underbrace{1,1,1,\ldots 1}_{m \ \  \rm{times}} \ \underbrace{-1,-1, -1, \ldots -1}_{L-m\ \  \mathrm{times}}\}^T 
\end{equation}
and $\xi$ is a generic permutation, corresponding to a global phase transformation in QM. A full account of this will be given in a future paper describing RaQM in detail. 
 
 \end{document}